\documentclass[12pt]{article}

\usepackage{ol2}
\usepackage{amsmath}

\begin{document}

\title{Transient chirp in high speed photonic crystal quantum dots lasers with controlled spontaneous emission}

\author{R. Braive, S. Barbay, I. Sagnes, A. Miard, I. Robert-Philip and A. Beveratos}

\address{
Laboratoire de Photonique et Nanostructures LPN-CNRS UPR-20, Route de Nozay, 91460 Marcoussis, France\\
$^*$Corresponding author: Alexios.Beveratos@lpn.cnrs.fr }

\begin{abstract}
We report on a series of experiments on the dynamics of
spontaneous emission controlled nanolasers. The laser cavity is a
photonic crystal slab cavity, embedding self-assembled quantum
dots as gain material. The implementation of cavity electrodynamics
effects increases significantly the large signal modulation
bandwidth, with measured modulation speeds of the order of 10 GHz
while keeping an extinction ratio of 19 dB. A linear transient
wavelength shift is reported, corresponding to a chirp of less
than 100 pm for a 35-ps laser pulse. We observe that the chirp
characteristics are independent of the repetition rate of the
laser up to 10 GHz.
\end{abstract}

\ocis{140.3948; 270.5580; 140.5960; 250.5590; 350.4238}

\maketitle

Recent progress in the design and fabrication of microcavities
\cite{Song2005} has enabled to implement cavity quantum
electrodynamics (CQED) effects in solid state. In the weak
coupling regime, the acceleration and spatial redistribution of
spontaneous emission has recently been exploited in the
engineering of non-conventional lasers with high spontaneous
emission coupling factors $\beta$ \cite{Strauf2006, Ulrich2007}.
Such cavity enhanced lasers promise to have a large direct
intensity modulation bandwidth \cite{Yamamoto1991}. Large signal
modulation bandwidth up to 100 GHz has already been observed with
quantum wells photonic crystal lasers \cite{Atlug2006}, although
lasing in such structures is strongly affected by high non-radiative
recombinations rates with an increase of the laser threshold
power\cite{Englund2007}. Quantum dots are less sensitive to free
surface non-radiative traps thanks to the three-dimensional
confinement of carriers. In quantum dots CQED-enhanced lasers,
modulation speeds up to 30 GHz are expected \cite{Ellis2007}. Yet,
a major problem in direct modulation semiconductor lasers is the
large change of carrier density during pulse emission leading to a
detrimental large frequency chirp of the gain-switched pulses.
Therefore, it is highly desirable to fabricate a laser yielding
minimum chirp for high speed operation. In this paper, by studying
time-resolved spectra and the spectral-resolved temporal evolution
of a CQED-enhanced photonic crystal laser embedding quantum dots,
we demonstrate that large direct modulation bandwidth up to 10 GHz
can be achieved, with a linear spectral chirp of less than 100 pm
within the 35 ps pulse width. Linear chirp could be compensated by standard
compensation techniques.

The laser cavity is formed by a photonic crystal double
heterostructure \cite{Song2005} etched on a 180 nm-thick suspended
GaAs membrane and incorporating a single layer of self-assembled
InAs quantum dots at its vertical center plane. The whole
structure is grown by molecular beam epitaxy. The quantum dot
density is of the order of 4.10$^{10}$ cm$^{-2}$ and their
spontaneous emission is centered around 945 nm at 4 K with an
inhomogeneous broadening of about 30 nm. The cavity is fabricated
using electron beam lithography, inductively coupled plasma
etching and wet etching \cite{Braive2008}. It consists of three
segments of W1 photonic crystal waveguides. The intermediate
segment extends over two periods with a longitudinal period
$a_c=250$ nm, that  is locally enhanced compared to the
longitudinal period $a_m=240$ nm of the two surrounding segments.
The targeted air hole-radius $r$ is 0.29$\times a_m$.

\begin{figure}[h!]
   \begin{center}
   \begin{tabular}{c}
   \includegraphics[width=8.3cm]{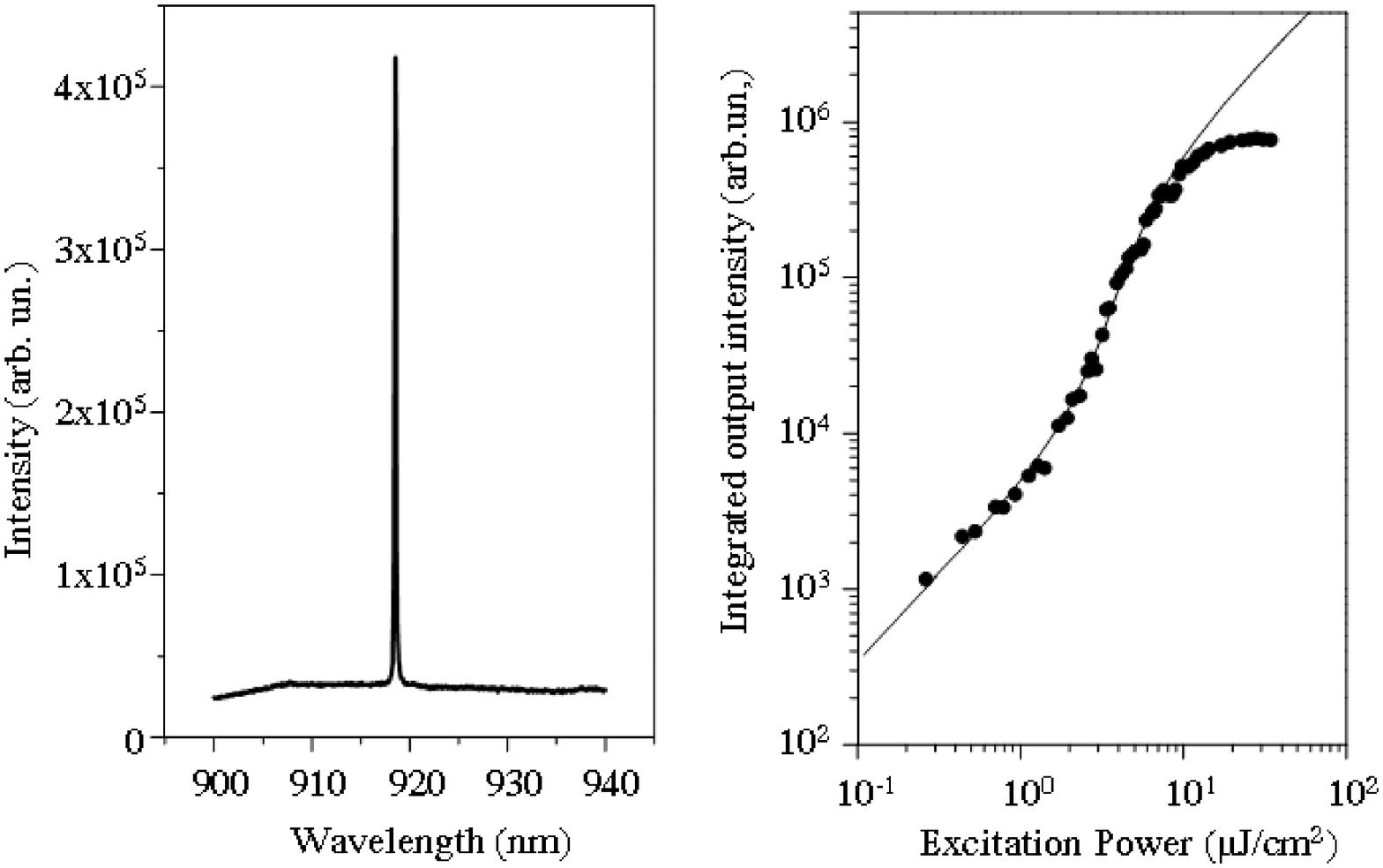}
   \end{tabular}
   \end{center}
   \caption[example]
   { \label{Fig1} Left: Laser spectrum of the cavity for an excitation power of 7.76 $\mu$J$/$cm$^{2}$ (1.7$\times P_{th}$) measured onto the surface sample. Right: Integrated output intensity as a function of the excitation power. Full circles: experimental data. Solid line: Rate equation model solution corresponding to $\beta$ of 0.67.}
   \end{figure}

\begin{figure}[h]
   \begin{center}
   \begin{tabular}{c}
   \includegraphics[width=8.3cm]{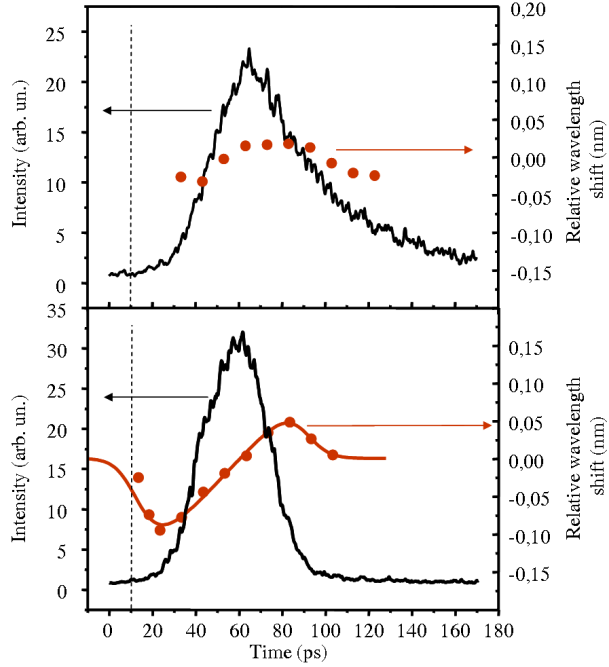}
   \end{tabular}
   \end{center}
   \caption{Time-resolved optical emission from the photonic crystal cavity. (Upper) below the laser threshold ($P\simeq 0.5 P_{th}=$2.1$\mu$J$/$cm$^2$), (Lower) above the laser threshold ($P\simeq 5P_{th}=$22$\mu$J$/$cm$^2$). The red circles represent the relative wavelength shift of the spectral maximum intensity as a function of time. Red line corresponds to a fit by Eq. \ref{Eq1}. Vertical dashed lines represent the temporal position of the laser excitation pulse.}
   \label{Fig2}
   \end{figure}

The cavity is cooled at 4 K. The active material is optically
and non-resonantly pumped at normal incidence by a Ti:Sapphire laser
delivering 3 ps pulses at a 81.8 MHz repetition rate and tuned to
840 nm near the energy gap of the wetting layer, in order to
reduce the impact of thermal heating of the membrane. The pumping
laser pulses are focused by a microscope objective (numerical
aperture=0.4) onto the sample. The photoluminescence emitted by
the cavity is collected through the same microscope objective and
sent on a 32 cm spectrometer ($\simeq$ 0.1 nm resolution) equipped
with a cooled silicon charge coupled device. The signal dispersed
by the spectrometer is also sent on a Syncroscan streak camera,
with a temporal resolution of 3 ps. A typical laser spectrum is
shown on Fig. \ref{Fig1}{a} and shows a sharp single-mode lasing
peak. Lasing occurs around 920 nm, on the short-wavelength side of
the quantum dot spectral distribution as observed in
\cite{Xie2007}. Cavity modes on resonance with the long-wavelength
side of the quantum dot spectral distribution do not display any
lasing behavior, since material gain saturation occurs before
laser threshold is reached. In order to evidence lasing operation
in such cavity, we recorded the integrated output intensity of the
cavity mode as a function of pump power. Figure \ref{Fig1}(b)
shows the resulting Light-in Light-out ($L-L$) curve. A s-shaped
smooth intensity transition appears at intermediate excitation
levels of the order of $P_{th}=$4.5 $\mu$J$/$cm$^2$. This soft
lasing transition has already been observed in CQED-enhanced
nanolaser \cite{Strauf2006, Ulrich2007}. Determination of the
spontaneous emission coupling factor $\beta$ in such cavities is
usually obtained from fits of the $L-L$ curve by coupled rate
equation models for carrier density $N$ and photon density in the
cavity $P$ \cite{Strauf2006}. Yet, such method leads to a certain
degree of uncertainty on the induced value of $\beta$, since these
equations involve a large number of parameters. Moreover, the
output characteristics of such cavity lasers are strongly
dependent on the excitation regime \cite{Gies2008}. For comparison purpose, $L-L$ curves are fitted to the rate equation model with standard values of
the linear gain and transparent carrier density as proposed in
\cite{Strauf2006}, we estimate a $\beta$ factor around factor $\beta=$0.67 depicted in solid line in Fig.
\ref{Fig1}{b}. We consider a quantum dot lifetime equal to $\tau_{sp}=50$ ps as measured below threshold (Fig. \ref{Fig2}a) corresponding to a Purcell factor of F$_p$=20 consolidating the high $\beta$ parameter. For
large pumping powers above 6.5 $\mu$J$/$cm$^2$, we observe a
deviation from the theoretical fit, due to gain saturation effects
that are not included in the model.

To investigate the dynamics of our laser, emission from the cavity
above and below threshold is analyzed on the streak camera.
Typical decay curves are shown on Fig. \ref{Fig2}. As expected in
CQED enhanced lasers and as already observed in \cite{Ellis2007},
we observe a decrease of the pulse rise time from values of the
order of 25 ps below threshold to values of the order to 11 ps for
pumping powers of 6.6 $\mu$J$/$cm$^2$ (of the order of
1.5$P_{th}$), since spontaneous emission rapidly builds up the
photon number in the cavity mode. For higher pumping powers ($>$
1.5$P_{th}$), this rise time is pinned to 11 ps, certainly limited
by the capture time. Simultaneously, as the pump rate increases,
we observe a strong decrease of the decay time, from 50 ps below
threshold to 11 ps above threshold. The minimum decay rate is
observed again at pump powers around 1.5$P_{th}$ and the decay
time remains unchanged for higher pumping rates. This excitation
power of 1.5$P_{th}$ corresponds to the excitation power at which
gain saturation effects appear (see Fig. \ref{Fig1} left). On
Figure \ref{Fig2}, are also reported the variations of the
emission wavelength within the pulse duration, deduced from traces
recorded on the streak camera. For low excitation powers below
threshold, the wavelength shift is rather small, less than 0.18
$\dot{A}$ within the spectral resolution of the setup. Conversely,
well above threshold, we observe a transient chirp, with first a
slight blueshift, followed by a continuous linear redshift and
then a blueshift. The linear redshift of 100 pm occurs over a
time scale of $\Delta \tau=35$ ps corresponding to the full width
at half maximum of the laser pulse. The blueshift results from
carrier-induced change of the refractive index, whereas the
redshift appears when lasing occurs, decreasing consequently
rapidly the total number of carriers. When modeling the delivered
temporal laser pulse by a Gaussian-shaped pulse in the form of
$e^{-(\frac{t}{\Delta\tau/(2\sqrt{ln2})})^2}$, the laser time-dependent
wavelength chirp associated with the dynamic change of the carrier
density can be described under large signal modulation by
\cite{Koyama1985}:
\begin {equation}
\delta \lambda(t) =  \alpha\frac{\lambda^2}{2\pi c}
(\frac{t}{(\Delta\tau/(2\sqrt{ln2}))^2} +\frac{\beta
N_{th}}{2\tau_{sp}P}.e^{(\frac{t}{\Delta\tau/(2\sqrt{ln2})})^2})
\label{Eq1}
\end{equation}
where  $\alpha$ is the linewidth enhancement factor. $N_{th}$ is
the carrier density at threshold, $P$ is the maximum photon density in the
cavity and $\tau_{sp}$ is the spontaneous emission lifetime. When
fitting our curve by Eq. \ref{Eq1}, the time evolution of the
chirp and the small total chirp are well reproduced, using a
fitting parameter $N_{th}/P$ of 0.17 and $\beta=0.67$. The value of $\alpha$,
equal to 3.05 in this fit, is deduced from time-bandwidth product
measurements : $\Delta \nu \Delta \tau = \frac{2ln2}{\pi}
\sqrt{1+\alpha^2}$, where $\Delta \nu$ is the full spectral width
at half maximum. Such measurements repeated for different
excitation powers indicate that $\alpha$ enhances with the pump
power \cite{Melnik2006} from 2.62 up to 3.05, when the excitation
power raises from $P_{th}$ to 5$P_{th}$. This increase results
mainly from the increase of $\Delta\nu$, since $\Delta \tau$ is
roughly constant around 35 ps. This enhancement of $\alpha$ may
result from gain material saturation effects, inducing low
differential gains and thus higher $\alpha$ factors
\cite{Muszalski2004}. This $\alpha$ factor value, of the order of
the ones observed on quantum well lasers, is large compared to
theoretical predictions for quantum dots material gain but rather
small compared to the values obtained on conventional quantum dots
lasers operating above threshold \cite{Markus2003}.

\begin{figure}[h]
   \begin{center}
   \includegraphics[width=8.3cm]{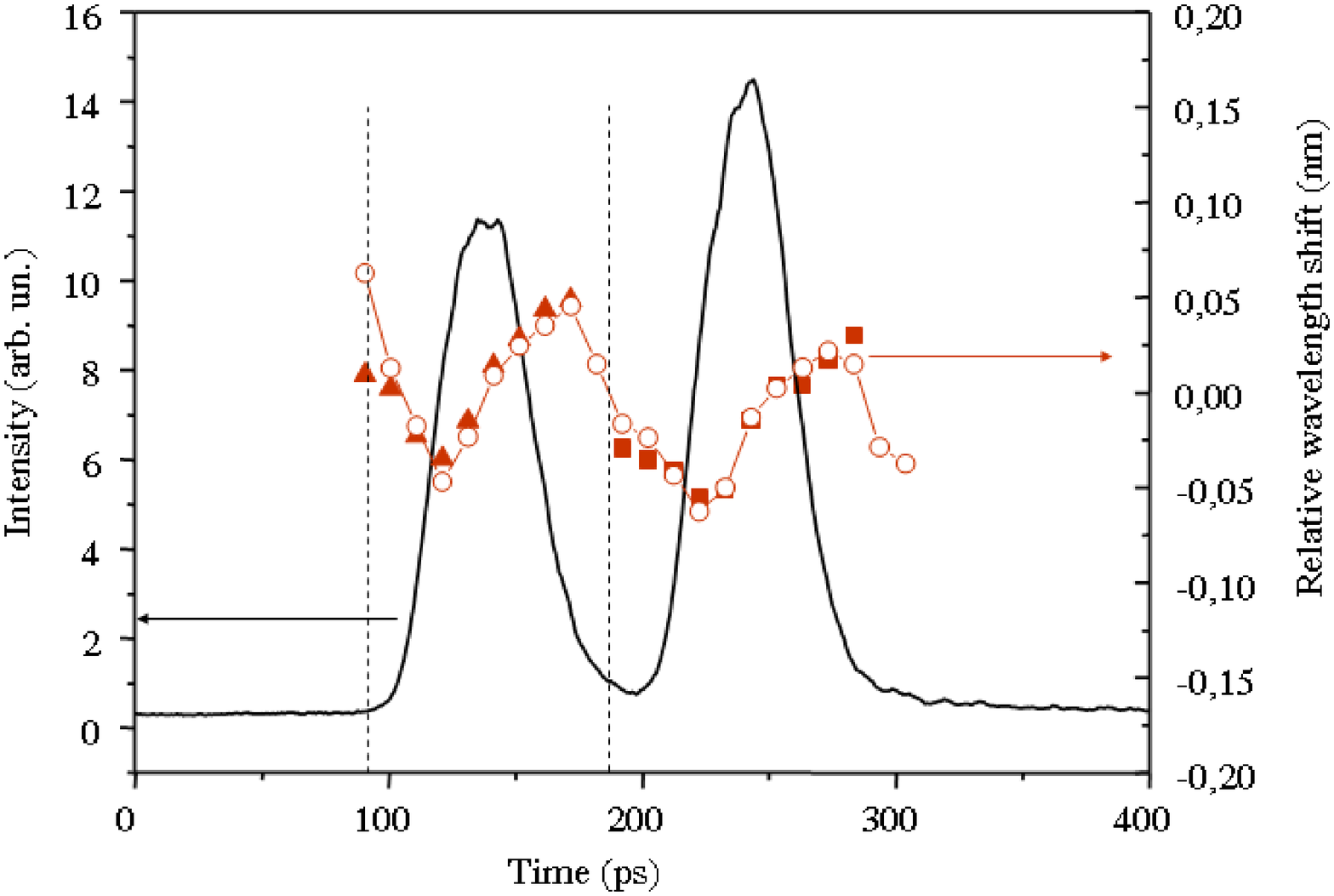}
   \end{center}
   \caption{Temporal and spectral response of the laser cavity excited by two 3-ps pulses separated by 100 ps. The temporal position of the excitation pulses are indicated by vertical dash lines. The excitation power is set to 5$P_{th}$. The chirp has been measured, when the device is excited by only the first (full triangles), only the second (full squares) or both excitation pulses (circles). }
   \label{Fig3}
\end{figure}

When applying a direct large-signal modulation to the device, it
is important that all pulses delivered at high repetition rate
present the same linear wavelength dependency, allowing for an
effective group velocity dispersion compensation independently of
the modulation pattern. Figure \ref{Fig3} presents the temporal
response of our laser, when excited by two subsequent 3-ps pulses
separated by 100 ps. The response of the laser follows the pump
with a clear 19 dB extinction rate between the two pulses: the
photonic crystal laser delivers two identical 34 ps width pulses
with a time difference of 100 ps. The spectral variation of the
optical frequency has also been measured, when the sample is
pumped by only one of the two excitation pulses and when the
sample is pumped by the two subsequent excitation pulses. We
observe that the chirp is identical in the two-pulse excitation
regime to that observed in the one-pulse excitation regime for
each pulse. Moreover, in the two-pulse excitation regime, the
variations of the wavelength during the pulses duration, are
identical in both pulses. This result indicates that the whole
system has relaxed on a time scale smaller than 100 ps, since the
characteristics of the second pulse are not affected by the first
one.

In summary, the spectral response in time of modulated CQED
enhanced lasers at large-signal direct modulation rates of 10 GHz
has been investigated. Our results indicate a small linear
transient chirp within each pulse duration, that could be
compensated by use of highly dispersive single-mode fibres. The
chirp characteristics of subsequent delivered pulses do not depend
on the direct modulation bandwidth up to 10 GHz. These results
confirm that quantum dot photonic crystal lasers with high
spontaneous emission coupling factors may improve significantly
technologies for high pulse repetition rate applications such as
optical interconnections.

The authors would like to thank A. Lema\^itre for
samples growth as well as R. Kuszelewicz and P. Voisin for useful
discussions. The authors also thank K. Meunier and J. Bloch for
lending critical electronic equipment.

\end{document}